\documentclass[fleqn,usenatbib]{mnras}

\usepackage{newtxtext,newtxmath}

\usepackage[T1]{fontenc}
\usepackage{times}

\usepackage{graphicx}	
\usepackage{amsmath}	
\usepackage{amssymb}	
\usepackage{siunitx}
\usepackage[ruled]{algorithm2e}
\usepackage{float}

\newcommand{\dd}{\mathrm{d}} 

\newcommand{\fuv}{f_\text{\tiny{UV}}}
\newcommand{\ms}{\text{M}_\odot}
\newcommand{\arad}{\textbf{a}_\text{rad}}
\newcommand{\agrav}{\textbf{a}_\text{grav}}
\newcommand{\atot}{\textbf{a}_\text{tot}}
\newcommand{\taux}{\tau_\text{\tiny X}}
\newcommand{\tauuv}{\tau_\text{\tiny UV}}

\title[Qwind code release]{Qwind code release: a non-hydrodynamical approach to modelling line-driven winds in active galactic nuclei}

\author[Arnau Quera-Bofarull et al.]{Arnau Quera-Bofarull,$^{1,2}$\thanks{E-mail: arnau.quera-bofaurll@durham.ac.uk}
Chris Done$^{3}$,
Cedric Lacey$^{1}$,
Jonathan C. McDowell$^{4}$,
\newauthor
Guido Risaliti$^{5,6}$,
Martin Elvis$^{4}$
\\
$^{1}$Institute for Computational Cosmology, Department of Physics, Durham University, South Road, Durham DH1 3LE, UK\\
$^{2}$Institute for Data Science, Durham University, South Road, Durham DH1 3LE, UK\\
$^{3}$Centre for Extragalactic Astronomy, Department of Physics, University of Durham, South Road, Durham DH1 3LE, UK\\
$^{4}$Harvard-Smithsonian Center for Astrophysics, 60 Garden St, Cambridge, MA 02138, USA\\
$^{5}$Dipartimento di Fisica e Astronomia, Universit\`a degli Studi di Firenze, Via. G. Sansone 1,50019 Sesto Fiorentino (FI), Italy.\\
$^{6}$INAF - Osservatorio Astrofisico di Arcetri, Largo E. Fermi 5 50125 Firenze, Italy.
}

\date{Accepted XXX. Received YYY; in original form ZZZ}

\pubyear{2019}

\begin{document}
\label{firstpage}
\pagerange{\pageref{firstpage}--\pageref{lastpage}}
\maketitle

\begin{abstract}
Ultraviolet (UV) line driven winds may be an important part of the active galactic nucleus (AGN) feedback process, but understanding their impact is hindered by the complex nature of the radiation hydrodynamics. 
Instead, we have taken the approach pioneered by Risaliti \& Elvis, calculating only ballistic trajectories from radiation forces and gravity, but neglecting gas pressure. We have completely re-written their {\sc Qwind} code using more robust algorithms, and can now quickly model the acceleration phase of these winds for any AGN spectral energy distribution spanning UV and X-ray wavebands. We demonstrate the code using an AGN with black hole mass $10^8\, \ms$ emitting at half the Eddington rate and show that this can effectively eject a wind with velocities $\simeq (0.1-0.2)\, c$. The mass loss rates can be up to $\simeq 0.3 M_\odot$ per year,
consistent with more computationally expensive hydrodynamical simulations, though we highlight the importance of future improvements in radiation transfer along the multiple different lines of sight illuminating the wind. The code is fully public, and can be used to quickly explore the conditions under which AGN feedback can be dominated by accretion disc winds.
\end{abstract}




\section{Introduction}

Almost every galaxy in the Universe hosts a supermassive black hole (BH) at its centre. It is observationally well grounded that the BH mass ($M_\text{BH}$) correlates with different galactic-scale properties such as the bulge's stellar mass \citep{haring_black_2004} and velocity dispersion \citep{ferrarese_fundamental_2000, gebhardt_relationship_2000-1}
which suggests a joint evolution of the BH and its host galaxy \citep{magorrian_demography_1998, kormendy_coevolution_2013}. Nonetheless, the nature of the physical coupling between the BH and its host galaxy is not entirely understood, though winds from the accretion discs of supermassive black holes are a strong candidate to explain how the accretion energy can be communicated to much larger galactic scales. Observations show that (10-20)\% of quasars (QSOs) exhibit broad blueshifted absorption lines (BALs) with velocities of $ \varv \sim (0.03 - 0.3) \, c$ \citep{weymann_comparisons_1991-1, pounds_high-velocity_2003-1, pounds_evidence_2003, reeves_compton-thick_2009,  crenshaw_feedback_2012-1, tombesi_evidence_2010}. Many physical mechanisms have been proposed to explain the launching and acceleration phases of these outflows. Magnetic fields control the accretion process of the disc through the magnetorotational instability \citep{balbus_instability_1998, ji_hydrodynamic_2006}, enabling the transport of angular momentum outwards. It is therefore possible that they also play a key role in generating disc winds \citep{proga_numerical_2003, fukumura_magnetic_2017}, as well as being responsible for the production of radio jets \citep{blandford_electromagnetic_1977, blandford_hydromagnetic_1982}. Another plausible force that can accelerate a disc wind is radiation pressure onto spectral lines. The ultraviolet (UV) luminosity from the accretion disc can resonantly interact with the disc's surface gas through bound-bound line transitions, effectively boosting the radiative opacity by several orders of magnitude with respect to electron scattering alone, provided that the material is not overionised \citep[][hereafter SK90]{stevens_x-ray_1990}. This acceleration mechanism is also strongly supported by the observation of line-locking phenomena \citep{bowler_line-driven_2014}.

The physical principles of radiatively line-driven winds were extensively studied by \cite{castor_radiation-driven_1975}, hereafter CAK, and \cite{abbott_theory_1982} in the context of O-type stars. Two decades later the same approach was extended to accretion discs around active galactic nuclei (AGN) \citep{murray_accretion_1995}, using the classical thin disc model of \citet{shakura_black_1973} (hereafter SS). A few years later, the first results of hydrodynamical simulations of line-driven winds using the ZEUS2D code \citep{stone_zeus-2d_1992} were released \citep[][hereafter P00 and P04]{proga_dynamics_2000, proga_dynamics_2004}, and continue to be extensively improved \citep[][hereafter N16]{nomura_radiation_2016}, and also \cite{nomura_line-driven_2017, nomura_line-driven_2018, dyda_non-axisymmetric_2018, dyda_non-axisymmetric_2018-1}. 

However, full radiation hydrodynamic calculations are very computationally intensive. Another approach is to study only ballistic trajectories, i.e. 
neglect the gas pressure forces. This non-hydrodyamic approach was started by \cite{risaliti_non-hydrodynamical_2010-1}, hereafter RE10, as the radiation force from efficient UV line driving can be much stronger than pressure forces. 
Their \textsc{Qwind} code calculated the ballistic trajectories of material from an accretion disc illuminated by both UV and X-ray flux. The neglect of hydrodynamics means that the code can be used to quickly explore the wind properties across a wide parameter space, showing where a wind can be successfully launched and accelerated to the escape velocity and beyond. 

Here we revisit the \textsc{Qwind} code approach, porting it from C to Python, and improving it for better numerical stability and correcting some bugs. We show that this non-hydrodynamic approach does give similar results to a full hydrodynamic simulation. We illustrate how this can be used to build a predictive model of AGN wind feedback by showing the wind mass loss rate and kinetic luminosity for a typical quasar.
The new code, \textsc{Qwind2}, is now available as a public release on GitHub \footnote{\url{https://www.github.com/arnauqb/qwind}}. 

\section{Methods}
\label{sec:methods}

In this section we include for completeness the physical basis of the code
and its approach to calculating trajectories of illuminated gas parcels (RE10).
In subsection \ref{sec:geometry_setup} we describe the geometrical setup of the system. The treatment of the X-ray and UV radiation field is explained in subsection \ref{sec:radiation_field}, and we conclude by presenting the trajectory evolution algorithm in subsection \ref{sec:trajectory_evolution}. 

\subsection{Geometry setup}
\label{sec:geometry_setup}

We use cylindrical coordinates $(R, \phi, z)$, with the black hole and the X-ray emitting source considered as a point located at the centre of the grid, at $R = z = 0$. The disc is assumed to 
emit as a Novikov-Thorne \citep{novikov_astrophysics_1973} (NT) disc, but is assumed to be geometrically razor thin, placed in the plane $z=0$, with its inner radius given by $R_\text{isco}$ and outer radius at $R_\text{out}$. We model the wind as a set of streamlines originating from the surface of the disc
between radii $R_\text{in}\ge R_\text{isco}$ and $R_\text{out}$, where the freedom to choose $R_{in}$ allows wind production from the very inner disc to be suppressed by the unknown physical structure which gives rise to the X-ray emission.

The trajectory of a gas element belonging to a particular streamline is computed by solving its equation of motion given by $\textbf{a} = \textbf{f}_\text{grav} + \textbf{f}_\text{rad}$, where $\textbf{a}$ is the acceleration and $\textbf{f}_\text{grav}$ and $\textbf{f}_\text{rad}$ are the force per unit mass due to gravity and radiation pressure respectively, using a time-adaptive implicit differential equation system solver (sec. \ref{sec:trajectory_evolution}). The computation of the trajectory stops when the fluid element falls back to the disc or it reaches its terminal velocity, escaping the system. Since the disc is axisymmetric, it is enough to consider streamlines originating at the $\phi = 0$ disc slice.

\subsection{Radiation field}
\label{sec:radiation_field}

The radiation field consists of two spectral components. 

\subsubsection{The X-ray component}
    The central X-ray source is assumed to be point-like, isotropic, and is solely responsible for the ionisation structure of the disc's atmosphere. The X-ray luminosity, $L_\text{\tiny{X}} = f_\text{\tiny{X}} \, L_\text{bol}$. The ionisation parameter is
    \begin{equation}\label{eq:ionisation_parameter}
        \xi = \frac{4 \pi F_\text{\tiny{X}}}{n},
    \end{equation}
where $F_\text{\tiny{X}}$ is the ionising radiation flux, and $n$ is the number density. The X-ray flux at the position $(R,z)$ is computed as
\begin{equation}\label{eq:xray_flux}
    F_\text{\tiny{X}} = \frac{L_\text{\tiny{X}} \; \exp{(-\tau_\text{\tiny{X}}})}{4 \pi r^2},
\end{equation}
where $r = \sqrt{R^2 + z^2}$, and $\tau_\text{\tiny{X}}$ is the X-ray optical depth, which is calculated from 
\begin{equation}\label{eq:tau_x}
    \tau_\text{\tiny{X}} = \int_{R_\text{in}}^r \; n_0(r') \; \sigma_\text{\tiny{X}} (\xi)\; \dd r',
\end{equation}
where $n_0(r)$ is the number density measured at the base of the wind, by projecting the distance $r'$ to the disc (see Appendix \ref{app:optical_depths}). This assumption overestimates both the UV and the X-ray optical depth far from the disc surface. Since most of the acceleration is gained very close to the disc, the impact of this assumption is small, however, it will be improved in a future improvement of the radiative transfer model. $\sigma_{\tiny X}(\xi)$ is the cross-section to X-rays as a function of ionisation parameter, which we parametrise following the standard approximation from \cite{proga_dynamics_2000},
\begin{equation}
    \sigma_\text{\tiny{X}}(\xi) = \begin{cases} 100 \, \sigma_\mathrm{\tiny{T}}&\text{ if } \xi < 10^5 \text{ erg cm s}^{-1}, \\ \sigma_\mathrm{\tiny{T}} &\text{ if } \xi \geq 10^5 \, \text{ erg cm s}^{-1},
    \end{cases}
\end{equation}
where the step function increase in opacity below $\xi = 10^5 \text{ erg cm s}^{-1}$ very approximately accounts for the increase in opacity due to the bound electrons in the inner shells of metal ions, and $\sigma_\mathrm{\tiny T}$ is the Thomson cross section. 

\subsubsection{The ultraviolet component}
The UV source is the accretion disc, emitting according to the NT model in an anisotropic way due to the disc geometry. The UV luminosity is $L_\text{\tiny{UV}} = f_\text{\tiny{UV}} \, L_\text{bol}$. Currently the code makes the simplifying assumption that $f_\text{\tiny{UV}}$ is constant as a function of radius. The emitted UV radiated power per unit area by a disc patch located at $(R_d, \phi_d, 0)$ is
\begin{equation}
\mathcal F = \fuv\frac{3G M\dot M}{8 \pi R_d^3} f(R_d, R_\text{isco}).
\end{equation}
The SS equations as used by RE10 are non-relativistic, with 
$f(R_d,R_\text{isco})=[1-(R_\text{isco}/R_d)^{1/2}]$ which leads to the standard Newtonian disc bolometric luminosity of $L_d=\frac{1}{12}\dot{M}c^2$ i.e. an efficiency of $\approx 0.08$ for a Schwarzschild black hole, with $R_\text{isco}=6R_g$. We use instead the fully relativistic NT emissivity, where $f$ is explicitly a function of black hole spin, $a$, and the efficiency is the correct value of $\eta(a=0)=0.057$ for a Schwarzschild black hole. This is important, as the standard input parameter, $\dot{m}=L_{\rm bol}/L_{\rm Edd}$, is used to set $\dot{M}$ via $L_{\rm bol}/(\eta(a) c^2)$. The relativistic correction reduces the radiative power of the disc by up to 50\% in the innermost disc annuli, compared to the Newtonian case. 

Assuming that the radiative intensity (energy flux per solid angle) $I(R_d)$ is independent of the polar angle over the range $\theta \in [0, \pi / 2]$, we can write 
\begin{equation}
    I(R_d) = \frac{\mathcal F}{\pi},
\end{equation}
thus the UV radiative flux from the disc patch as seen by a gas blob at a position $(R,0,z)$ is  
\begin{equation}
    \dd F = \fuv\frac{I(R_d)}{\Delta^2}\; \cos \theta \;R_d\; \dd R_d \dd\phi_d,
\end{equation}
where
\begin{equation}
    \Delta = (R^2 + R_d^2 + z^2 - 2 \, R \, R_d \cos \phi_d)^{1/2}.
\end{equation}
(The flux received from an element of area $\dd A =R_d\; \dd R_d \dd\phi_d$ at distance $\Delta$ seen at angle $\theta$ is $I \dd\Omega$, where the solid angle subtended is $\dd\Omega = (\dd A \,\cos \theta)  /\Delta^2$, and $\cos \theta = z / \Delta$.)

The average luminosity weighted distance is $\Delta \approx r$, so attenuation by electron scattering along all the UV lines of sight is approximately that along the line of sight to the centre i.e. analogously to equation (\ref{eq:tau_x}), but only considering the electron scattering cross-section (see Appendix \ref{app:optical_depths}). A more refined treatment that considers the full geometry of the disc will be presented in a future paper. 
The corresponding radiative acceleration due to electron scattering is then
\begin{equation}
    \dd \, \arad = \frac{\sigma_\mathrm{\tiny{T}}}{c}\; \hat{\textbf{n}} \; \dd F \; \exp{(-\tau_\text{\tiny{UV}})} \,,
\end{equation}
with $\hat{\textbf{n}}$ being the unit vector from the disc patch to the gas blob,
\begin{equation}
    \hat{\mathbf{n}} = \frac{( R - R_d \cos \phi_d, -R_d \sin \phi_d, z)}{\Delta}.
\end{equation}

\subsubsection{Radiative line acceleration}
\begin{figure}
    \centering
    \includegraphics[width=\columnwidth]{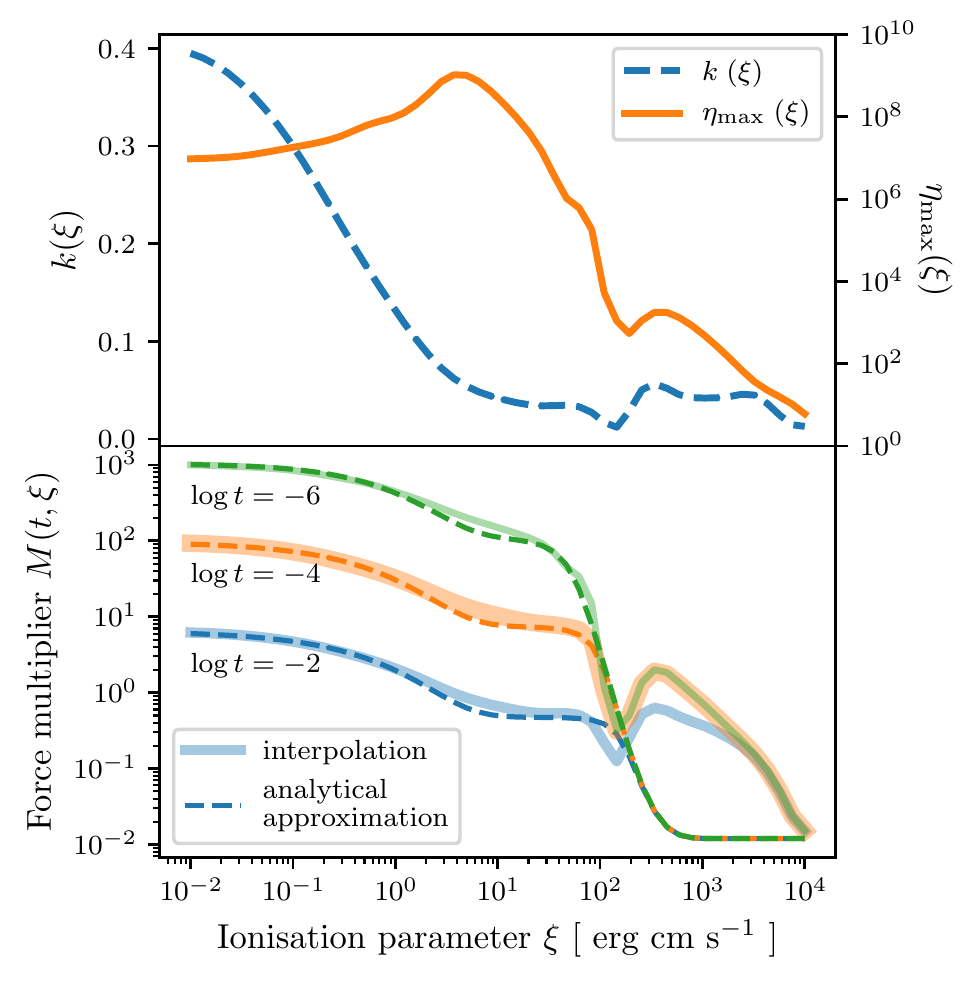}
    \caption{Top panel: Best fit values for the force multiplier parameters $k$ and $\eta_\mathrm{max}$ as a function of ionisation parameter $\xi$, taken from SK90. Bottom panel: force multiplier as a function of the ionisation parameter and the effective optical depth, showing the discrepancy between the analytical approximation derived in SK90 and the direct interpolation at the range $10^2 \leq \xi \leq 10^4$. Note that, for the analytical approximation, $M$ is independent of $t$ for $\xi > 10^2$.}
    \label{fig:force_multiplier}
\end{figure}

The full cross-section for UV photons interacting with a moderately ionised gas is dominated by line absorption processes, implying potential boosts of up to 1000 times the radiation force caused solely by electron scattering. To compute this, we use the force multiplier $M$ proposed by \cite{stevens_x-ray_1990} hereafter SK90, which is a modified version of \cite{castor_radiation-driven_1975}  that includes the effects of X-ray ionisation. Ideally, one should recompute the force multiplier considering the full AGN Spectral Energy Distribution (SED) (\cite{danne2019b}), which is different than the B0 star spectrum considered in SK90, however this is out of the scope of this paper. The full opacity is then $\sigma_\text{total} = (1+M) \; \sigma_\text{\tiny T}$, with the force multiplier $M$ depending on the ionisation parameter, and on the effective optical depth parameter $t$,
\begin{equation}
\label{eq:tau_eff}
    t = \sigma_\text{\tiny T} \; n \;\varv_\text{th} \; \left| \frac{\dd \varv}{\dd l}\right|^{-1},
\end{equation}
which takes into account the Doppler shifting resonant effects in the accelerating wind, and depends on the gas number density $n$, the gas thermal velocity $\varv_\text{th}$ and the  spatial velocity gradient along the light ray, $\dd\varv / \dd l$. In general, the spatial velocity gradient is a function of the velocity shear tensor and the direction of the incoming light ray at the current point. In this work we approximate the velocity gradient as the gradient along the gas element trajectory, allowing the force multiplier to be determined locally. A full velocity gradient treatment in the context of hydrodynamical simulations of line driven winds in CV systems has been studied in \cite{dyda_non-axisymmetric_2018}, who find that the inclusion of non-spherically symmetric terms results in the formation of clumps in the wind. Our non-hydrodynamical approach is insensitive to this kind of gas feature. It is convenient to rewrite the spatial velocity gradient as
\begin{equation}
\label{eq:dv_dr}
    \frac{\dd \varv}{\dd l} = \frac{\dd \varv}{\dd t} \; \frac{\dd t}{\dd l} = \frac{a_t}{v_t},
\end{equation}
where $a_t = \sqrt{a_R^2 + a_z^2}$, and $v_t = \sqrt{\varv_R^2 + \varv_z^2}$. This change of variables avoids numerical roundoff errors as it avoids calculating small finite velocity differences.
The force multiplier is parametrised as 
\begin{equation}\label{eq:force_multiplier}
    M(t, \xi) = k(\xi) \, t^{-0.6} \left[ \frac{(1 + t\;\eta_\text{max}(\xi))^{0.4} - 1}{(t\;\eta_\text{max}(\xi))^{0.4}}\right] \approx k(\xi) \, t^{-0.6},
\end{equation}
where the latter expression holds when $\eta_\text{max}(\xi) \, t \gg 1$, which is the case for all cases of interest here. We extract the best fit values for $k$ and $\eta_\text{max}$ directly from Figure 5 of SK90, as opposed to using the usual analytic approximation given in equations 18 and 19 of SK90. The reason we fit directly is because the analytical fitting underestimates the force multiplier in the range $10^2 \leq \xi \leq 10^4$, as we can see in Figure \ref{fig:force_multiplier}. In RE10 the analytical approximation was used, but we note that the step function change in X-ray opacity at $\xi=10^5$ means that these intermediate ionisation states are not important in the current handling of radiation transfer, since the gas quickly shifts from being very ionised to being neutral, thus this change has negligible effect on the code results.

With all this in mind, the total differential radiative acceleration is 
\begin{equation}
    \dd \,\arad = \frac{\sigma_\text{\tiny T}\; ( 1 + M(t, \xi))}{c}\; \hat{\textbf{n}} \; \dd F \; \exp{(-\tau_\text{\tiny{UV}})}\,,
\end{equation}
and the contribution from the whole disc to the radial and vertical radiation force is found by performing the two integrals
\begin{equation}\label{eq:Ir}
    \mathcal I_R  =\int_{R_\text{min}}^{R_\text{max}}\, \dd R_d \, \frac{f(R_d, R_\text{isco})}{R_d^2} \int_0^\pi \dd \phi_d \frac{(R - R_d \cos \phi_d)}{\Delta^4},
\end{equation}
and
\begin{equation}\label{eq:Iz}
    \mathcal I_z  = \int_{R_\text{min}}^{R_\text{max}} \, \dd R_d  \, \frac{f(R_d, R_\text{isco})}{R_d^2} \int_0^\pi \dd \phi_d \, \frac{1}{\Delta^4}.
\end{equation}
The angular contribution is zero because of the cylindrical symmetry. Evaluating these integrals is not straightforward due to the presence of poles at $\Delta = 0$. The original \textsc{Qwind} code used a fixed grid spacing, but this is not very efficient, and led to inaccuracies with convergence of the integral (see section \ref{sec:baseline}). Instead, we use the {\sc Quad} integration method implemented in the Scipy \citep{virtanen_scipy_2019} Python package 
to compute them. Appendix \ref{app:integrals} shows that this 
converges correctly. 

\subsection{Trajectories of fluid elements}
\label{sec:trajectory_evolution}

Gas trajectories are initialised at a height $z_0$, with launch velocity 
$\varv_0$. This can be different to the assumed thermal velocity as there could be additional mechanisms which help launch the wind from the disc, such as convection and/or magnetic fields, thus we keep this as a free parameter in the code so we can explore the effect of this. The equation of motion is $\atot = \agrav + \arad$, with
\begin{equation}
\label{eq:gravity}
    \agrav\,(R,z) = -\frac{GM_\mathrm{BH}}{r^2}\,\left(\frac{R}{r},\, 0,\, \frac{z}{r}\right).
\end{equation}
In cylindrical coordinates, the system to solve is
\begin{equation}
\label{eq:trajectory_ode}
    \begin{split}
        &\frac{\dd R}{\dd t} - \varv_R &= 0,\\
        &\frac{\dd z}{\dd t} - \varv_z &= 0,\\
        &\frac{\dd \varv_R}{\dd t} - a^\mathrm{grav}_R - a^\mathrm{rad}_R - \frac{\ell^2}{R^3} &= 0,\\
        &\frac{\dd \varv_z}{\dd t} - a^\mathrm{grav}_z - a^\mathrm{rad}_z&=0,\\
    \end{split}
\end{equation}
where $\ell$ is the specific angular momentum, which is conserved along a trajectory. The radiative acceleration depends on the total acceleration and the velocity at the evaluating point through the force multiplier (see equations (\ref{eq:force_multiplier}) and (\ref{eq:dv_dr})), therefore, the system of differential equations cannot be written in a explicit form, and we need to solve the more general problem of having an implicit differential algebraic equation (DAE), $\textbf{F}(t, \textbf{x}, \dot{\textbf{x}}) = 0$, where $\textbf{F}$ is the LHS of equation (\ref{eq:trajectory_ode}), $\textbf{x} = (R, z, \varv_R, \varv_z)$, and $\dot{\textbf{x}}=(\varv_R, \varv_z, a_R, a_z)$. We use the IDA solver \citep{hindmarsh_sundials_2004} implemented in the \textsc{Assimulo} simulation software package \citep{andersson_assimulo_2015}, which includes the backward differentiation formula (BDF) and an adaptive step size to numerically integrate the DAE system. We choose a BDF of order 3, with a relative tolerance of $10^{-4}$. In RE10, a second order Euler method was used without an adaptive time step. We do not find significant differences in the solutions found by both solvers, as RE10 used a very small step size, keeping the algorithm accurate. Nonetheless, the time step adaptiveness of our new approach reduces the required number of time steps by up to 4 orders of magnitude, making the algorithm substantially faster. For an assessment on the solver's convergence refer to Appendix \ref{app:integrals}.

The gas density is calculated using the mass continuity equation, $\dot M_\text{line}(t) = \dot M_\text{line}(0)$. If the considered streamline has an initial width $\Delta L_0$, assuming that the width changes proportionally to the distance from the origin, $\Delta L \propto r$, we can write
\begin{equation}
    \label{eq:mass_loss}
    \dot M_0 = \rho_0 \, \varv_0 \, A_0 = \rho_0 \, \varv_0 \, 2 \, \pi \, r_0 \, \Delta L_0 = \rho \, \varv \, 2 \, \pi \, r \, \Delta L = \dot M,
\end{equation}
where $\rho_i = n(r_i)\, m_p$ with $m_p$ being the proton mass. From here, it easily follows, using $\Delta L / \Delta L_0 = r / r_0$, that
\begin{equation}
    \label{eq:update_density}
    n(r_{i})\; \varv_i \; r_i^2 = n(r_0)\; \varv_0 \; r_0^2,
\end{equation}
which we use to update the density at each time step.
The simulation stops either when the fluid element falls back to the disc, or when it leaves the grid ($r=10^5 \; R_g$). 

\section{The \textsc{Qwind2} code}
\label{sec:code_structure}

\begin{algorithm}
\SetAlgoLined
\SetKwInOut{Input}{input}
\Input{$R_0$, $z_0$, $n_0$, $\varv_0$}
Read initial parameters\;
Set initial angular velocity to Keplerian\;
Initialise IDA solver\;
\While{(material not out of grid) or (material not fallen to the disc)}
{
    IDA solver iteration. At each step, take current value of $\textbf{x}$, and $\dot{\textbf{x}}$, and do:
    
    \Indp
    Compute local velocity gradient $\frac{\dd \varv}{\dd l}$\ using (\ref{eq:dv_dr})\;
    Compute gas density using (\ref{eq:update_density})\;
    Compute X-ray and UV optical depth (see Appendix \ref{app:optical_depths})\;
    Compute ionisation parameter using (\ref{eq:ionisation_parameter}) and (\ref{eq:xray_flux})\;
    Compute force multiplier using (\ref{eq:force_multiplier})\;
    Compute radiative acceleration using the computed force multiplier and integrating equations (\ref{eq:Iz}) and (\ref{eq:Ir})\;
    Compute gravitational acceleration using (\ref{eq:gravity})\;
    
    Update fluid element position, velocity, and acceleration\;
    Estimate solver error and update time step\;
}
\uIf{gas escaped}{
    Compute mass loss using mass flux conservation (\ref{eq:mass_loss_formula})\;
    Compute kinetic luminosity using (\ref{eq:kin_lumin})\;
}{}
\caption{Fluid element trajectory initialisation and evolution}
\label{alg:1}
\end{algorithm}
\begin{figure}
    \centering
    \includegraphics[width=\columnwidth]{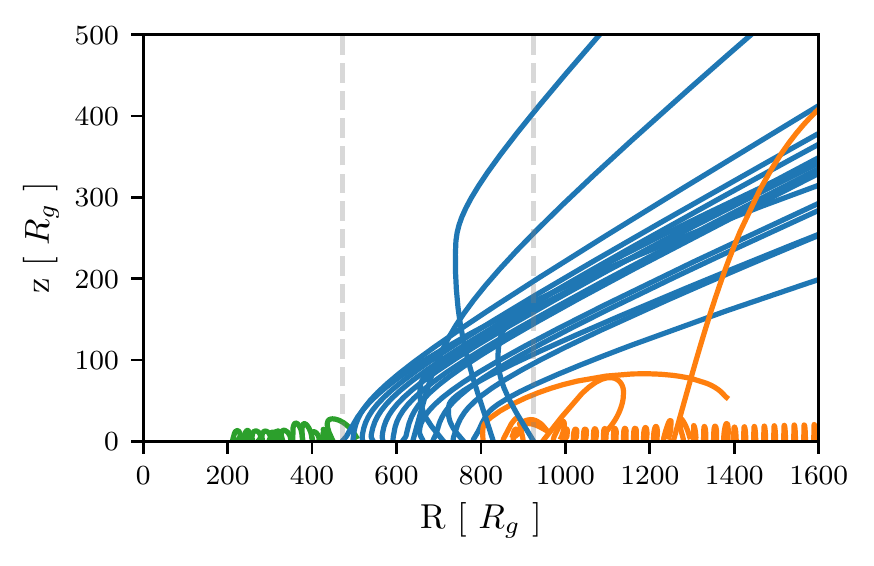}
    \caption{Results of a wind simulation, using the original \textsc{Qwind} code with the Newtonian disc flux equations from SS and a radiative efficiency of $\eta=0.0313$. The wind temperature is set to $T=2\times 10^6$ K. The other parameters are set to the baseline parameter values (Table \ref{table:baseline}). The inner failed wind, escaping wind, and outer failed wind regions are coloured in green, blue, and orange respectively, and delimited by the grey dashed vertical lines.}
    \label{fig:example}
\end{figure}{}

In the code, we organise the different physical phenomena into three Python classes: \emph{wind}, \emph{radiation}, and \emph{streamline}. The \emph{wind} class is the main class of the code and it handles all the global properties of the accretion disc and launch region, such as accretion rate, atmospheric temperature/velocity/density etc. The \emph{radiation} class implements all the radiative physics, such as the calculation of optical depths and the radiation force.
Finally, the \emph{streamline} class represents a single fluid element, and it contains the \textsc{Assimulo}'s IDA solver that solves the fluid element equation of motion, evolving it until it falls back to the disc or it exceeds a distance of $r=10^5 \; R_g$. It takes about 10 seconds on average on a single CPU to calculate one fluid element trajectory, thus we are able to simulate an entire wind in a few minutes, depending on the number of streamlines wanted.

The system is initialised with the input parameters (see Table \ref{table:baseline}), and a set of fluid elements are launched and evolved between $R_\text{in}$ and $R_\text{out}$ following Algorithm \ref{alg:1}. As an illustrative example, we define our baseline model with the parameter values described in Table \ref{table:baseline}. These parameter values are the same as used in RE10, except for the black hole mass that we take to be $M=10^8 \; \ms$, rather than $2 \, \times \, 10^8 \; \ms$, to be able to compare with the hydrodynamic simulations of P04 and N16. We also launch the wind from closer to the disc, at $z_0=1R_g$ rather than the default $z_0=5R_s=10R_g$ of RE10. We do this to highlight the effect of the new integration routine. 

To determine the number of streamlines $N$ to simulate, we notice that the mass flow along a streamline with initial radius $R_0$ is 
\begin{equation}
    \dot M_\text{wind} = 2 \pi\; R_0\; \rho_0 \; v_0 \; \Delta L_0,
\end{equation}
where $\Delta L_0 = (R_\text{out} - R_\text{in}) / N$. The streamline with the highest mass flow is, thus, the one with the highest initial radius. We expect that the outermost escaping streamline will satisfy, at its base, $\tau \simeq R_0 \; n_0\; \sigma_T \simeq 1$. We set $N$ such that this streamline carries, at most, the $0.05\%$ of the mass accretion rate. This implies that the chosen number of streamlines is independent of the initial density,
\begin{equation}
\label{eq:n_streamlines}
    N = \frac{2\pi \; m_\text{p} \; \varv_0 \; (R_\text{out} - R_\text{in})}{5 \times 10^{-4} \; \dot M \; \sigma_T}.
\end{equation}
For the parameter values of the baseline model (Table \ref{table:baseline}), we have $N\simeq 53$.

\begin{table}
\centering
\caption{Qwind baseline parameters.}
\begin{tabular}[t]{ll}
\hline
Parameter & Value\\
\hline
$R_\text{in}$& 200\, $R_g$\\
$R_\text{out}$& 1600\, $R_g$\\
$M_\text{BH}$ & $10^8\; M_\odot$ \\
$\dot m$& 0.5 \\
$a$ & 0 \\
$\varv_0$ & $10^7 \mathrm{cm s}^{-1}$\\
$n_0$ & $2\times 10^8$ cm$^{-3}$\\
$z_0$ & 1 $R_g$ \\
$T$ & $2.5\times 10^4$ K\\
$f_\text{\tiny{UV}}$ & 0.85\\
$f_\text{\tiny{X}}$ & 0.15\\
\hline
\label{table:baseline}
\end{tabular}
\end{table}%
\subsection{Improvements in the \textsc{Qwind} code}

\begin{figure*}
\centering
\includegraphics[width=\textwidth]{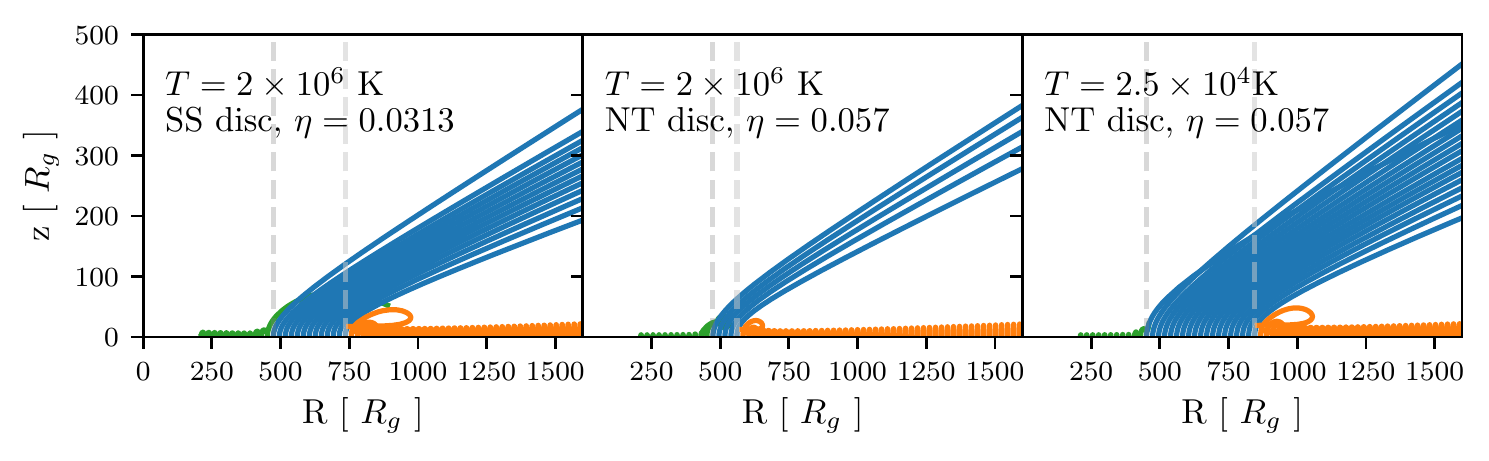}
\caption{Three wind simulations performed with the updated Qwind code but different physical assumptions. All the non specified parameters are fixed to the baseline model (Table \ref{table:baseline}). The leftmost panel shows a simulation run assuming a non-relativistic Newtonian disc with an efficiency of $\eta=0.0313$, and a temperature of $T=2\times 10^6 K$, while the middle one corresponds to the relativistic NT disc model with the correct efficiency $\eta = 0.057$. Finally, for the rightmost panel we use the relativistic disc model with the correct efficiency, and a temperature of $T=2.5\times 10^4$ K, consistent with the force multiplier derivation in SK90.}
\label{fig:baseline_comparison}
\end{figure*}

We first run the original \textsc{Qwind} code using the SS disc model with an efficiency $\eta=0.0313$, and a wind temperature of $T=2\times 10^6 K$. The rest of the parameters are fixed to the default values shown in Table \ref{table:baseline}. We plot the resulting streamlines in Figure \ref{fig:example}. The structure of the wind can be divided into three distinct regions: an inner failed wind (green), an escaping wind (blue), and an outer failed wind (orange), also delimited by the separating vertical dashed lines.
The inner failed region corresponds to streamlines which have copious UV irradiation but where the material is too highly ionised for the radiation force to counter gravity. On the other hand, the outer failed wind comprises trajectories where the material has low enough ionisation for a large force multiplier, but the UV flux is not sufficient to provide enough radiative acceleration for the material to escape. Finally, the escaping wind region consists of streamlines where the material can escape as it is shielded from the full ionising flux by the failed wind in the inner region.

\subsubsection{Effect of integration routine}

Two of the blue escaping wind streamlines in Figure \ref{fig:example} (those originating from $\sim 900 R_g$) cross all the other escaping trajectories. We find that these crossing flowlines result from the old integration routine. 
The original code solved the integrals (\ref{eq:Ir}) and (\ref{eq:Iz}) using a non-adaptive method, which led to numerical errors in the radiative force at low heights. 
The first panel of Figure \ref{fig:baseline_comparison} shows the results using the same parameters and code with the new integration routine. The behaviour is now much smoother, not just in the escaping wind section but across all of the surface of the disc. The new Python integrator is much more robust, and has much better defined convergence (see Appendix \ref{app:integrals}).

\subsubsection{Efficiency and disc emissivity}

The original code used the Newtonian disc flux equations from SS, but then converted from $\dot{m}=L_{bol}/L_{Edd}$ to $\dot M$ using an assumed efficiency, with default of $\eta = 0.0313$. 
This is low compared to that expected for the Newtonian SS disc accretion, where $\eta=0.08$, and low even compared to a fully relativistic non-spinning black hole which has $\eta = 0.057$. For a fixed dimensionless mass accretion rate $\dot m$, the inferred $\dot{M}\propto 1/\eta$ as a larger mass accretion rate is required to make the same bolometric luminosity if the efficiency is smaller. Since $\dot{M}$ sets the local flux, this means that the local flux is a factor of $\sim 2$ smaller in the new \textsc{Qwind2} code for a given $L_{bol}$. The comparison between the first and second panels of Figure \ref{fig:baseline_comparison} shows that this reduction in the local UV flux means that fewer wind streamlines escape. 

\subsubsection{Wind thermal velocity}

In the absence of an X-ray ionising source, the force multiplier is independent of the thermal velocity (see \cite{abbott_theory_1982} for a detailed discussion), this, however, does not mean one can freely choose a thermal velocity value at which to evaluate the effective optical depth $t$, since the values of the fit parameters in the analytical fit for $M(t)$ depend on the thermal velocity as well (see Table 2 in \cite{abbott_theory_1982}). If one includes an X-ray source as in SK90, then we expect the temperature of the gas to change depending on how much radiation a gas element is receiving from the X-ray source as well as from the UV source. At low values of the ionisation parameter, the results from \cite{abbott_theory_1982} hold, and thus it is justified to evaluate the force multiplier using an effective temperature of $T=2.5 \times 10^4 \; K$, corresponding to the temperature of the UV source used in SK90. Since the evaluation of the force multiplier is most important in regions of the flow where the gas is shielded from the X-ray radiation, we use this temperature value to compute the force multiplier throughout the code. In the N16 and P04 hydrodynamic simulations, the wind kinetic temperature is calculated by solving the energy equation that takes into account radiative cooling and heating, however, for the purpose of evaluating the force multiplier, a constant temperature of $T=2.5 \times 10^4 \; K$ is also assumed. In RE10, the force multiplier is evaluated setting $T=2\times 10^6$ K for the kinetic temperature. A higher thermal velocity increases the effective optical depth $t$, which in turn decreases the force multiplier given the same spatial velocity gradient and assuming the same parametrisation for $k(\xi)$ and $\eta_\text{max}(\xi)$, thus resulting on a narrower range of escaping streamlines. We can visualise the impact of this change by comparing the second and third panels of Figure \ref{fig:baseline_comparison}. However, this apparent dependence of the force multiplier on thermal velocity is artificial, as explained above.

We define our baseline model as the one with the parameters shown in Table \ref{table:baseline}.

\subsection{Baseline model in \textsc{Qwind2}}
\label{sec:baseline}

\begin{figure}
    \centering
    \includegraphics[width = \columnwidth]{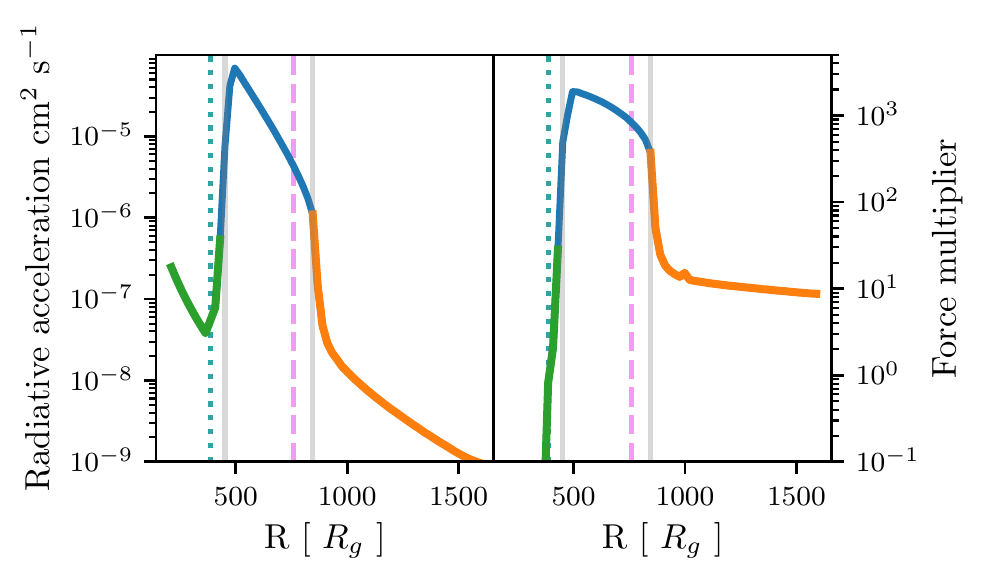}
    \caption{Maximum radiative force and force multiplier as a function of the initial radius of each streamline. Note that escaping lines, plotted in blue and delimited by the two vertical grey lines, require a balance between a sufficiently high force multiplier (thus low ionisation parameter), and high radiative force. Gas trajectories originating at the green coloured radii (left region delimited by the first vertical grey line) are too ionised, while the orange ones (rightmost region delimited by the second grey line) intercept too few UV photons. The radius at which the gas on the base of the wind becomes optically thick ($\tau = 1$) to X-Rays and UV is denoted by the dotted blue and the dashed purple lines respectively.}
    \label{fig:baseline_fm_force}
\end{figure}

\begin{figure}
    \centering
    \includegraphics[width = \columnwidth]{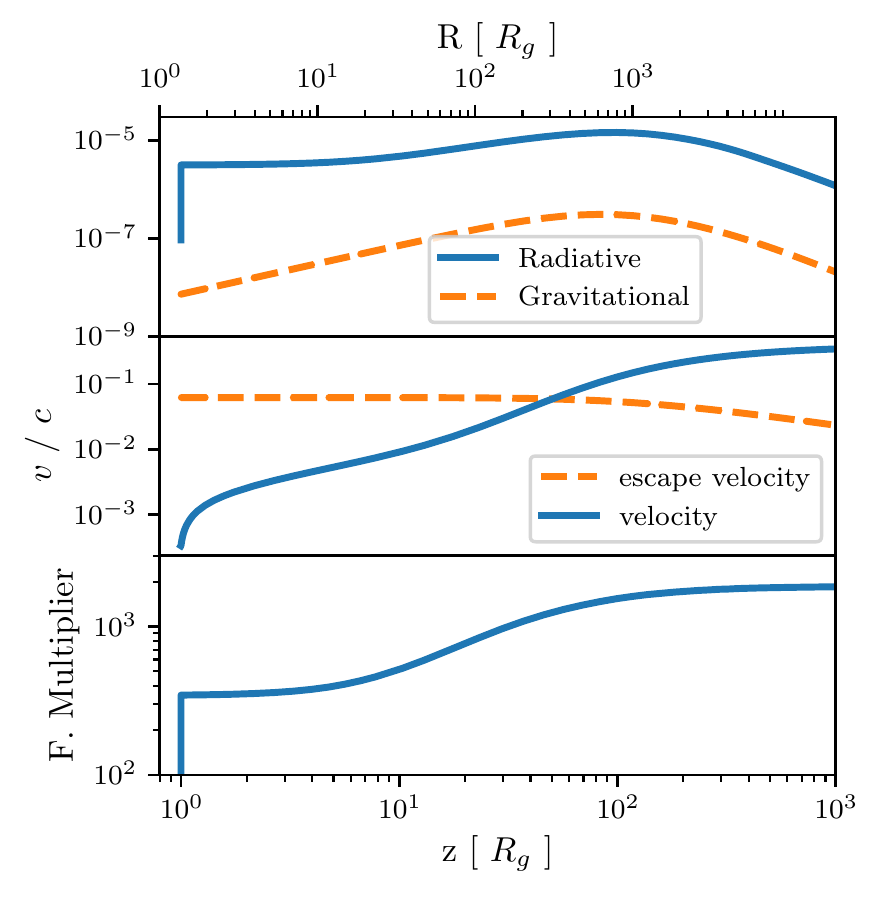}
    \caption{Streamline properties for an escaping gas trajectory. Top panel: Vertical radiative and gravitational acceleration as a function of height and radius. Middle panel: Streamline velocity as a function of radius and height. Bottom panel: Force multiplier as a function of radius and height.}
    \label{fig:streamline_individual}
\end{figure}

\begin{figure}
    \centering
    \includegraphics[width=\columnwidth]{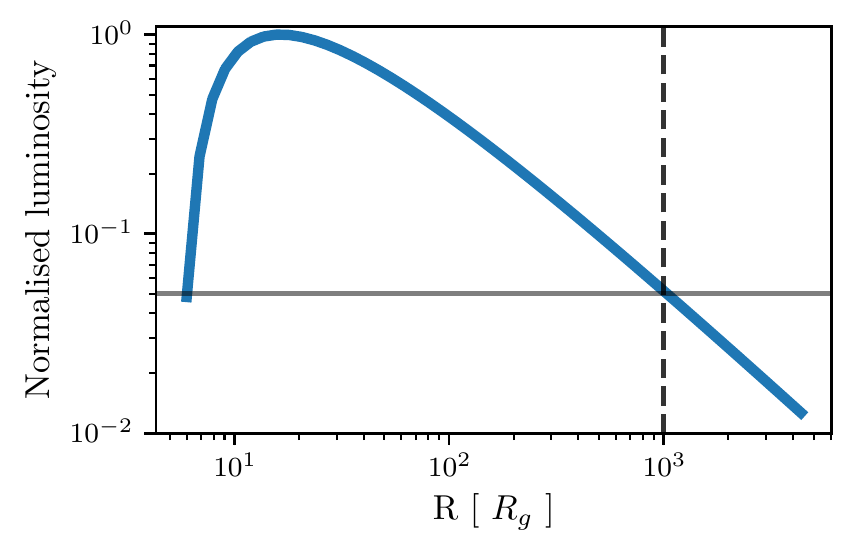}
    \caption{Disc annulus luminosity as a function of annulus radius, normalised to the luminosity of the brightest annulus. We have divided the radius range into 50 logarithmically spaced bins. The dashed black line corresponds to $R \simeq 1000 \; R_g$, from where the outer annuli contribute less than 5\% to the total luminosity compared to the brightest annulus at $R\simeq 16 R_g$. The sudden drop at $R \lesssim 16 R_g$ is due to the relativistic NT corrections to the SS disc.}
    \label{fig:radial_luminosity}
\end{figure}

\begin{figure*}
    \centering
    \includegraphics[width = \textwidth]{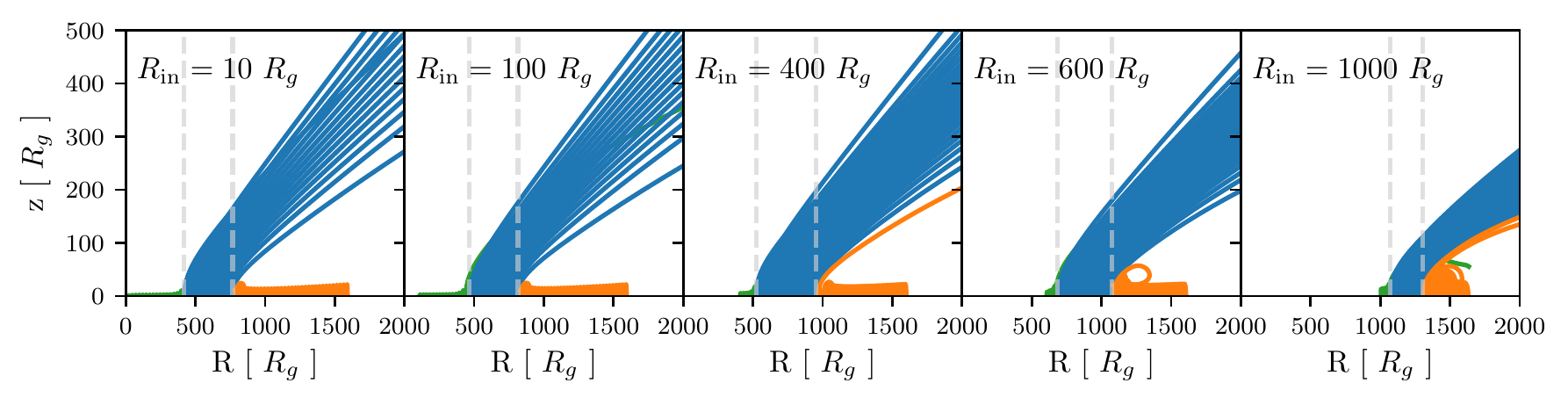}
    \caption{Different runs of the baseline model changing $R_\mathrm{in}$. The number of escaping streamlines is higher for larger values of $R_\mathrm{in}$, as the UV optical depth is lower while the shielding is still effective. Furthermore, outer escaping streamlines contribute more to the overall mass loss rate than the inner ones, since they represent larger disc annuli. The wind diminishes at $R_\mathrm{in} \gtrsim 1000 \; R_g$, where the disc annuli do not emit enough UV radiation.}
    \label{fig:rin_comparison}
\end{figure*}

The new code is publicly available online in the author's GitHub account \footnote{\url{https://github.com/arnauqb/qwind}}. It is written purely in Python, making use of the Numba \citep{lam_numba_2015} JIT compiler to speed up the expensive integration calculations. 

We now show more results from our new implementation of the \textsc{Qwind} code. The third panel of Figure \ref{fig:baseline_comparison} shows that the radius range from which escaping lines can be originated is relatively narrow. This can be explained by looking at the radiative acceleration and the force multiplier for each streamline. We plot the maximum radiative acceleration and force multiplier for each of the streamlines as a function of their initial radius in the left panel of Figure \ref{fig:baseline_fm_force}. To effectively accelerate the wind, we need both a high UV flux, and a high force multiplier, which requires that the X-ray flux is sufficiently attenuated. Therefore, computing the UV and X-ray optical depths from the centre at the base of the wind can give us an estimate of the escaping region. Indeed, the cyan dotted line shows the radius at which the optical depth along the disc becomes unity for X-ray flux, while the purple dashed line shows the same for the UV flux. Clearly this defines the radii of the escaping streamlines, i.e. successful wind launching requires that the X-rays are attenuated but the UV is not. 

We focus now on the physical properties of an individual escaping streamline. In Figure \ref{fig:streamline_individual}, we plot the vertical radiative acceleration, the velocity, and the force multiplier of the streamline as a function of its height and radius. We observe that most of the acceleration is achieved very rapidly and very close to the disc, consequently, the wind becomes supersonic shortly after leaving the disc, thus justifying our non-hydrodynamical approach. The sub-sonic part of the wind is encapsulated in the wind initial conditions, and the subsequent evolution is little affected by the gas internal forces. As we are focusing on a escaping streamline, the ionisation parameter is low, thus $\eta_\mathrm{max}$ will be very high (see top panel Fig. \ref{fig:force_multiplier}), enabling us to write $M(t) \propto  t^{-0.6}$ (by taking the corresponding limit in eq. (\ref{eq:force_multiplier})). Additionally, since the motion of the gas element is mostly vertical at the beginning of the streamline, we have from the continuity equation (eq. (\ref{eq:update_density})) $n \propto \varv_t^{-1}$, which combined with eq. (\ref{eq:dv_dr}) gives 
\begin{equation}
    M(t)\propto t^{-0.6} \propto  \left(\frac{a_t}{n v_t}\right)^{0.6} \simeq a_t^{0.6}.
\end{equation}
Therefore, as the gas accelerates, the force multiplier increases as well, creating a resonant process that allows the force multiplier to reach values of a few hundred, accelerating the wind to velocities of $\varv \sim (0.1 - 0.2) \; c$. At around $z = 50\; R_g$, the gas element reaches the escape velocity at the corresponding radius, and it will then escape regardless of its future ionisation state.

We use mass conservation to calculate the total wind mass loss rate by summing the initial mass flux of the escaping trajectories,
\begin{equation}\label{eq:mass_loss_formula}
\begin{split}
    \dot M_\text{wind} &= \sum_{i \in \left \{\substack{\text{escaping} \\ \text{trajectories}}\right \}} \dot M_\text{wind}^{(i)} \\
    &= \sum_{i \in \left \{\substack{\text{escaping} \\ \text{trajectories}}\right \}} \rho_{i,0} \; \varv_\text{i,0} \; 2\; \pi  \; R_{i,0} \; \delta R_i,
\end{split}
\end{equation}
where $\delta \, R_{i,0} = R_{i+1,0} - R_i$. For the baseline model we obtain $\dot M_\text{wind} = 3.01 \times 10^{24} $ g s$^{-1}$ $ = 0.05 \, \mathrm{M}_\odot \, $ yr$^{-1}$, which equates to $2.5\%$ of the black hole mass accretion rate.  We can also compute the kinetic luminosity of the wind, 
\begin{equation}\label{eq:kin_lumin}
    L_\text{kin} = \frac{1}{2} \;\dot M_\text{wind} \;\varv_\text{wind}^2\,,
\end{equation}
where $\varv_\text{wind}$ is the wind terminal velocity, which we take as the velocity at the border of our grid, making sure that it has converged to the final value. The wind reaches a kinetic luminosity of $L_\text{kin} = 7.83 \times 10^{43}\; \text{erg} / \text{s}$, which equates to 0.62\% of the Eddington luminosity of the system. Both these results depend on the choice of the initial conditions for the wind. In the next section, we scan the parameter range to understand under which parameter values a wind successfully escapes the disc, and how powerful it can be. 

\subsection{Dependence on launch parameters: \texorpdfstring{$R_{\text{in}},\, n_0,\, v_0$}{Rin, n0, v0}}

\begin{figure*}
    \centering
    \includegraphics[width=\textwidth]{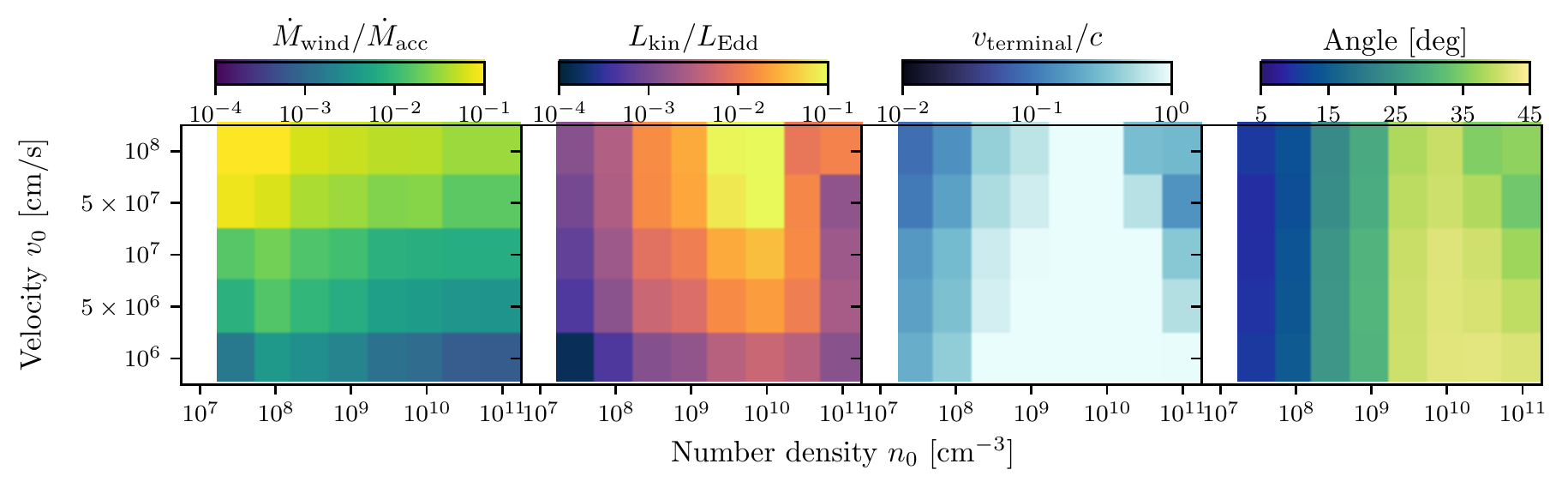}
    \caption{Results of different wind simulations varying the initial density $n_0$, and the initial velocity $\varv_0$. The rest of the parameters are fixed to the baseline model values (Table \ref{table:baseline}), except for $R_\text{in}=60\; R_g$, and $f_x=0.1$. The first panel shows the mass loss rate normalised to the mass accretion rate, and the second panel shows the wind kinetic luminosity normalised to the Eddington luminosity. Finally the third and fourth panels show the terminal velocity and angle of the fastest streamline in the wind.}
    \label{fig:init_conditions_scan_lowtemp}
\end{figure*}

\begin{figure*}
    \centering
    \includegraphics[width=\textwidth]{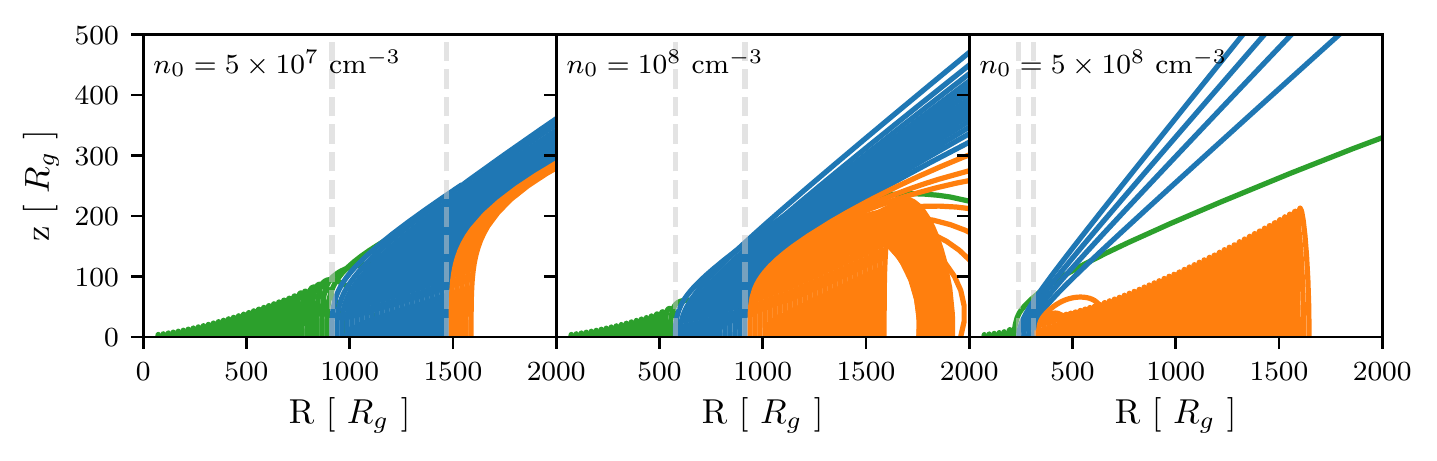}
    \caption{Three wind simulations from the parameter range scan shown in Figure \ref{fig:init_conditions_scan_lowtemp}. The initial velocity is fixed to $\varv_0 = 10^8$ cm/s, while the initial density is varied in the range $n_0 \in (5 \times 10^7,\; 10^8, \; 5 \times 10^8)$ cm$^{-3}$. Higher density values shift the escaping region to smaller radii, thus reducing the effective disc surface that produces an escaping wind. Furthermore, the increase in UV shielding narrows the range of escaping streamlines. Note that a high number of streamlines was used to produce Figure \ref{fig:init_conditions_scan_lowtemp}, according to the rule in equation (\ref{eq:n_streamlines}), but we plot only a fraction of them for clarity.}
    \label{fig:rho_comparison}
\end{figure*}

We consider variations around the baseline model (Table \ref{table:baseline}). We fix the black hole mass and accretion rate to their default values, and vary the initial launching radius $R_\text{in}$, the initial density $n_0$, and the initial velocity $\varv_0$. We can make some physical arguments to guide our exploration of the parameter space: 

\begin{enumerate}
    \item The initial radius $R_\text{in}$ at which we start launching gas elements can be constrained by considering the physical scale of the UV emitting region of the disc. In Figure \ref{fig:radial_luminosity}, we plot the luminosity of each disc annulus normalised to the luminosity of the brightest annulus, using 50 logarithmically spaced radial bins. We observe that radii larger than $\gtrsim 1000\; R_g$ contribute less than 5\% of the luminosity of the brightest annulus. On the other hand, the effective temperature of the disc drops very quickly below $R\simeq 16 \; R_g$ due to the NT relativistic corrections.We thus consider that the initial launching radius can vary from $10 \; R_g$ to $1000 \; R_g$.

In Figure \ref{fig:rin_comparison}, we plot the results of changing $R_\text{in}$ in the baseline model. Increasing the radius at which we start launching gas elements shifts the location of the wind towards higher radii, thus increasing the overall mass loss rate since outer streamlines represent a bigger disc surface (see equation (\ref{eq:mass_loss_formula})). For very large initial radii, $R_\text{in} \geq 1000 \; R_g$, the wind severely diminishes as the UV flux is too low.

To explore the remaining parameters, we fix $R_\text{in} = 60\; R_g$, and $f_x=0.1$. The reason for this is that we want to compare our results with the hydrodynamic simulations of P04 and N16, which used these parameter values. 
    
    \item The initial density $n_0$ of the gas elements needs to be high enough to shield the outer gas from the X-Ray radiation, so we need $\taux > 1$ at most a few hundred $R_g$ away from the centre (further away the UV flux would be too weak to push the wind). Therefore as a lower limit,
\begin{equation}
    \taux = \int_{R_\text{in}}^{100\; R_g} \, \sigma_x \, n_0 \, \dd r' < \int_{0}^{100\, R_g}\, 100\, \sigma_\text{\tiny T} \, n_0 \, \dd r' \simeq 10^{-7} \left(\frac{n_0}{\text{cm}^{-3}}\right),
\end{equation}
which implies a minimum shielding density of $n_0 \simeq 10^7 \; \text{cm}^{-3}$. On the other hand, if the density is too high the gas is also shielded from the UV flux coming from the disc. Even though our treatment of the UV optical depth assumes that the UV source is a central point source (see Appendix \ref{app:optical_depths}), let us consider now, as an optimistic case for the wind that the optical depth is computed from the disc patch located just below the wind. In that case, we need $\tauuv < 1$ at a minimum distance of $r \simeq 1\; R_g$,
\begin{equation}
    \tauuv = \int_0^{1R_g} \sigma_T \, n_0\, \dd r' \simeq 10^{-11} \left(\frac{n_0}{\text{cm}^{-3}}\right)\, ,
\end{equation}
so that the maximum allowed value is $n_0 \simeq 10^{11} \; \text{cm}^{-3}$. Thus, we vary the initial density from $10^7$ cm$^{-3}$ to $10^{11}$ cm$^{-3}$. 

\item Finally, we estimate the parameter range of the initial velocity $\varv_0$ by considering the isothermal sound speed at the surface of the disc. The disc's effective temperature at a distance of a few hundred $R_g$ from the centre computed with the NT disc model is $\simeq\, 10^6$ cm/s, so we vary the initial velocity from $10^6$ cm/s to $10^8$ cm/s to account for plausible boosts in velocity due to the launching mechanism. The total number of streamlines is adjusted to ensure enough resolution (see equation (\ref{eq:n_streamlines})). \end{enumerate}

Figure \ref{fig:init_conditions_scan_lowtemp} show the resulting scan over the
$n_0 - \varv_0$ parameter space.  These results confirm the physical intuition we described at the beginning of this section; initial density values lower than $\simeq 5 \times 10^7$ cm$^{-3}$ do not provide enough shielding against the X-ray radiation, while values higher than $\simeq 10^{10}$ cm$^{-3}$ shield the UV radiation as well, and produce a slower wind. Furthermore, lower initial velocities result into higher final velocities, as the gas parcels spend more time in the acceleration region, and are thus also launched at a higher angle with respect to the disc. The parameter combination that yields the highest wind mass loss rate is $n_0 = 5\times 10^7$ cm$^{-3}$ and $\varv_0=10^8$ cm/s, which predicts a mass loss rate of 0.3 $\ms$/yr, equal to $\simeq 17\%$ of the mass accretion rate. The reason why lower initial densities lead generally to higher mass loss rates can be visualised in Figure \ref{fig:rho_comparison}. Higher initial densities shift the wind launching region to the inner parts of the accretion disc, since they are able to shield the X-Ray more efficiently but the gas also becomes optically thick to UV radiation rapidly. On the other hand, for low values of the initial density, the gas becomes optically thick to X-Rays on the outer parts of the disc and the low UV attenuation implies that the range of escaping streamlines is wider. Additionally, outer radii represent annuli with bigger areas so the mass loss rate is significantly larger (see equation (\ref{eq:mass_loss_formula})). The parameter combination $n_0=10^{10}$ cm$^{-3}$ and $\varv_0=10^8$ cm/s yields the highest kinetic luminosity value,  however, a few of the escaping streamlines have non-physical superluminal velocities. The parameter combination that generates the physical wind with the highest kinetic luminosity is $n_0=5 \times 10^9$ cm$^{-3}$ and $\varv_0=10^8$ cm/s with $L_\text{kin} \simeq 9\% \, L_\text{Edd}$. Following \cite{hopkins_quasar_2010}, this kinetic energy would be powerful enough to provide an efficient mechanism of AGN feedback, as it is larger than $0.5\%$ of the bolometric luminosity. It is also worth noting that the angle that the wind forms with respect to the disc is proportional to the initial density. This can be easily understood, since, as we discussed before, higher initial densities shift the radii of the escaping streamlines inwards, from where most of the UV radiation flux originates. The wind originating from the inner regions of the disc has therefore a higher vertical acceleration, making the escaping angle higher compared to the wind in the outer regions.

\subsection{Comparison with hydrodynamic simulations}

A proper comparison with the hydrodynamic simulations of N16 and P04 is not straightforward to do, as there is not a direct correspondence of our free parameters with their boundary conditions, and some of the underlying physical assumptions are different (for instance, the treatment of the UV continuum opacity). Nonetheless, with P04 as reference, we have fixed so far $R_\text{in} = 60\, R_g$ to match their starting grid radius, and $f_x=0.1$, as they assume. 

\begin{figure}
    \centering
    \includegraphics[width=\columnwidth]{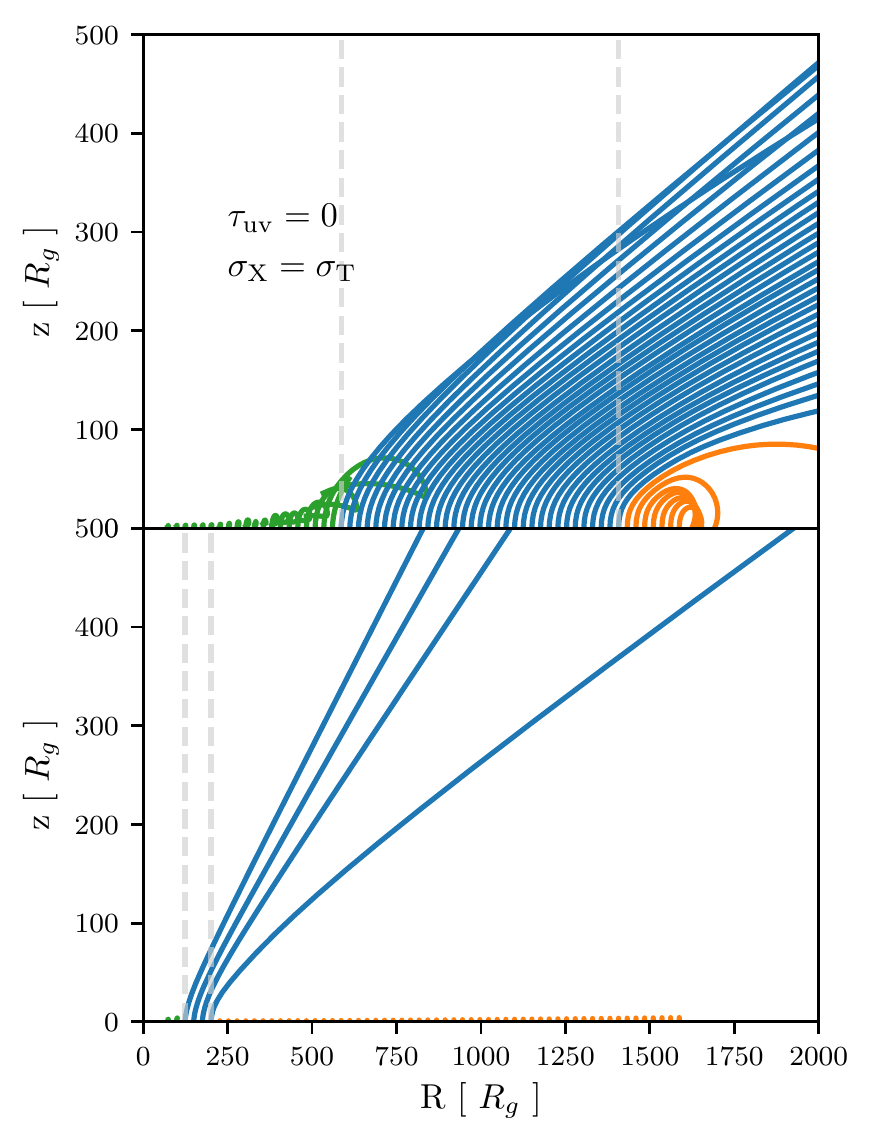}
    \caption{Top panel: Wind simulation with parameter values set to match P04: $f_x=0.1$, $R_\text{in}=60\; R_g$, $\varv_0=2\times10^6$ cm/s, $T=2.5\times 10^4 \;K$, and $n_0=2.5\times 10^9$ cm$^{-3}$. We also set $\tau_\text{\tiny UV} =0$, and $\sigma_\text{\tiny X} = \sigma_\text{\tiny T}$ as it is done in P04. Bottom panel: Wind simulation with same parameters as the top panel, but using the standard $\tau_\text{\tiny UV}$ and $\tau_\text{\tiny X}$ of \textsc{Qwind}}
    \label{fig:p04}
\end{figure}

Another physical assumption we need to change to compare with P04 is the treatment of the radiative transfer. In P04, the UV radiation field is not attenuated throughout the wind, although line self-shielding is taken into account by the effective optical depth parameter $t$. Furthermore, the X-ray radiation is considered to only be attenuated by electron scattering processes, without the opacity boost at $\xi \leq 10^5$ erg cm s$^{-1}$. We thus set $\tau_\text{\tiny UV} = 0$, and $\sigma_x = \sigma_\text{T}$. Finally, we assume that the initial velocity is $\varv_0=2 \times 10^6$ cm /s which is just supersonic at $T=2.5 \times 10^4$ K, and we fix $n_0 = 2.5 \times 10^9$ cm$^{-3}$, which gives $\tau_X=1$ at $r=100 \, R_g$. The result of this simulation is shown on the top panel Figure \ref{fig:p04}. We notice that not attenuating the UV continuum has a dramatic effect on the wind, allowing much more gas to escape as one would expect. Indeed, the bottom panel of Figure \ref{fig:p04} shows the same simulation but with the standard UV and X-ray continuum opacities used in $\textsc{Qwind}$. Running the simulation with the normal UV opacity but just electron scattering for the X-ray cross section results in no wind being produced. For the unobscured simulation that mimics P04, we obtain a wind mass loss rate of $0.3\, \ms \,/\, \text{yr}$, which is in good agreement with the results quoted in P04 ($\dot M_\text{wind} \sim (0.16 - 0.3)\, \ms \, / \, \text{yr}$). The wind has a kinematic luminosity of $L_\text{kin} = 0.7\%$ at the grid boundary, and a terminal velocity ranging  ($0.016 - 0.18$) c, again comparable to the range $(0.006 - 0.06)$ $c$ found in P04. Finally, the wind in P04 escapes the disc approximately at an angle between (4-21)$^{\circ}$, while in our case it flows at an angle in the range (3 - 14)$^{\circ}$.

{}

\section{Conclusions and future work}
\label{sec:conclusions}

We have presented an updated version of the \textsc{Qwind} code (\textsc{Qwind2}), aimed at modelling the acceleration phase of UV line-driven winds in AGNs. The consistency of our approach with other more sophisticated simulations shows that the non-hydrodynamical treatment is well justified, and that our model has the potential to mimic the results of more expensive hydrodynamical simulations.

The main free parameters of the model are the initial density and velocity of each streamline, and 
the inner disc radius from which the fluid elements are first launched. \cite{nomura_modeling_2013} calibrate the initial wind mass loss using the relation from CAK that links the wind mass loss from O-stars to their gravity and Eddington ratio. However, it is not clear whether this relation holds for accretion discs, where the geometry and the radiation field and sources are quite different \citep{laor_line_2014}. To be able to derive these initial wind conditions from first principles, we require a physical model of the vertical structure of the accretion disc. Furthermore, we need to take into account the nature of the different components of the AGN and their impact on the line-driving mechanism. In that regard, we can use spectral models like \cite{kubota_physical_2018-3} to link the initial conditions and physical properties of the wind to spectral features. We aim to present in an upcoming paper a consistent physical model of the vertical structure of the disc, considering the full extent of radiative opacities involved, that will allows us to infer the initial conditions of the wind. 

Another point that needs to be improved is the treatment of the radiation transfer. \textsc{Qwind} and current hydrodynamical simulations compress all of the information about the SED down to two numbers $L_\text{\tiny X}$ and $L_\text{\tiny UV}$, however, the wavelength dependent opacity can vary substantially across the whole spectrum. This simplification is likely to underestimate the level of ionisation of the wind \citep{higginbottom_line-driven_2014}, and motivates the coupling of \textsc{Qwind} to a detailed treatment of radiation transfer. \cite{higginbottom_simple_2013} construct a simple disc wind model with a Monte Carlo ionisation/radiative transfer code to calculate the ultraviolet spectra as a function of viewing angle, however, properties of the wind such as its mass flow rate and the initial radius of the escaping trajectories need to be assumed. We will incorporate a 
full radiative transfer code like $\textsc{Cloudy}$ or $\textsc{Xstar}$ to compute the line driving and transmitted spectra together. This also opens the possibility of having a metallicity dependent force multiplier, and studying how the wind changes with different ion populations.

Future development could also include dust opacity,
to study whether the presence of a dust driven wind can explain the origin of the broad line region in AGN \citep{czerny_origin_2011}.

The ability of \textsc{Qwind} to quickly predict a physically based wind mass loss rate make it very appealing to use as a subgrid model for AGN outflows in large scale cosmological simulations, as opposed to the more phenomenological prescriptions that are currently employed to describe AGN feedback.

\section*{Acknowledgements}
AQB acknowledges the support of STFC studentship (ST/P006744/1).
CD and CL acknowledge support by the the Science and Technology Facilities Council Consolidated Grant ST/P000541/1 to Durham Astronomy. 
This work used the DiRAC@Durham facility managed by the Institute for
Computational Cosmology on behalf of the STFC DiRAC HPC Facility
(\url{www.dirac.ac.uk}). The equipment was funded by BEIS capital funding
via STFC capital grants ST/K00042X/1, ST/P002293/1, ST/R002371/1 and
ST/S002502/1, Durham University and STFC operations grant
ST/R000832/1. DiRAC is part of the National e-Infrastructure.



\bibliographystyle{mnras}
\bibliography{qwind} 



\appendix

\section{Optical depth calculation}
\label{app:optical_depths}

\begin{figure}
    \centering
    \includegraphics[width=\columnwidth]{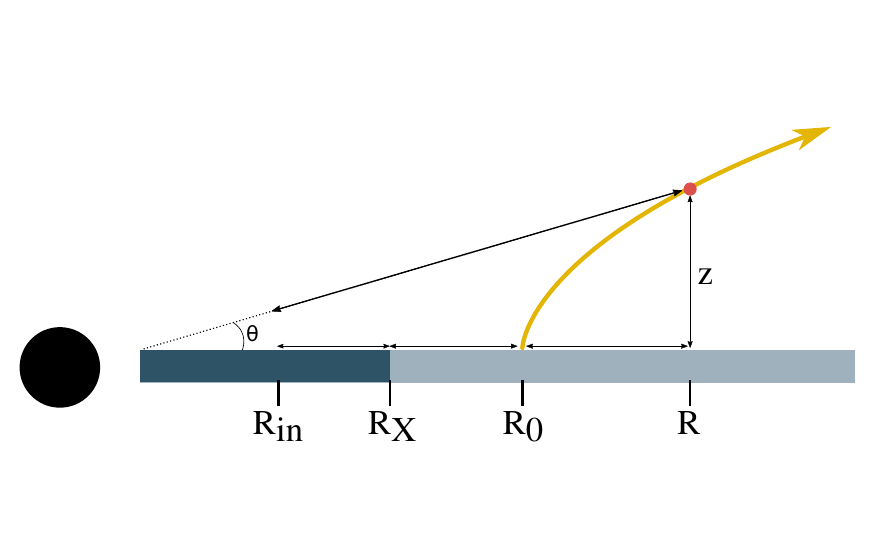}
    \caption{Schematic representation of the geometrical setup to compute the X-ray and UV optical depths. $R_0$ corresponds to the initial radius of the streamline being considered, and $R_x$ is the radius at which $\xi = 10^5$ erg cm s$^{-1}$. }
    \label{fig:xray_tau}
\end{figure}

The computation of the X-ray (and UV analogously) optical depth (eq. (\ref{eq:tau_x})) is not straightforward, as we need to take into account at which point the drop in the ionisation parameter boosts the X-ray opacity. Furthermore, the density is not constant along the light ray. Following the scheme illustrated in Figure \ref{fig:xray_tau}, $R_\mathrm{X}$ denotes the radius at which the ionisation parameter drops below $10^5 \,\mathrm{erg}\,\mathrm{cm}\, \mathrm{s}^{-1}$, $R_\text{in}$ is the radius at which we start the first streamline, and thus the radius from which the shielding starts, and finally $R_0$ is the initial radius of the considered streamline. With this notation in mind, we approximate the optical depth by
\begin{equation}
    \tau_\mathrm{X} = \sec \theta \,  \sigma_\mathrm{T} \, \left [ n_0 \, \int_{R_\text{in}}^{R_0} \kappa (R') \, \dd R' + n(R) \, \int_{R_0}^R \kappa(R')\, \dd R'\right],
\end{equation}
with
\begin{equation}
    \kappa(R) = \begin{cases} 100 &\text{ if } R > R_\mathrm{X}, \\ 1 &\text{ if } R \leq R_\mathrm{X}.
    \end{cases}
\end{equation}
The calculation for the UV optical depth is identical but setting the opacity boost factor to unity for all radii.

\section{Integral and solver convergence}
\label{app:integrals}

\subsection{Integral convergence}
\begin{figure}
    \centering
    \includegraphics[width=\columnwidth]{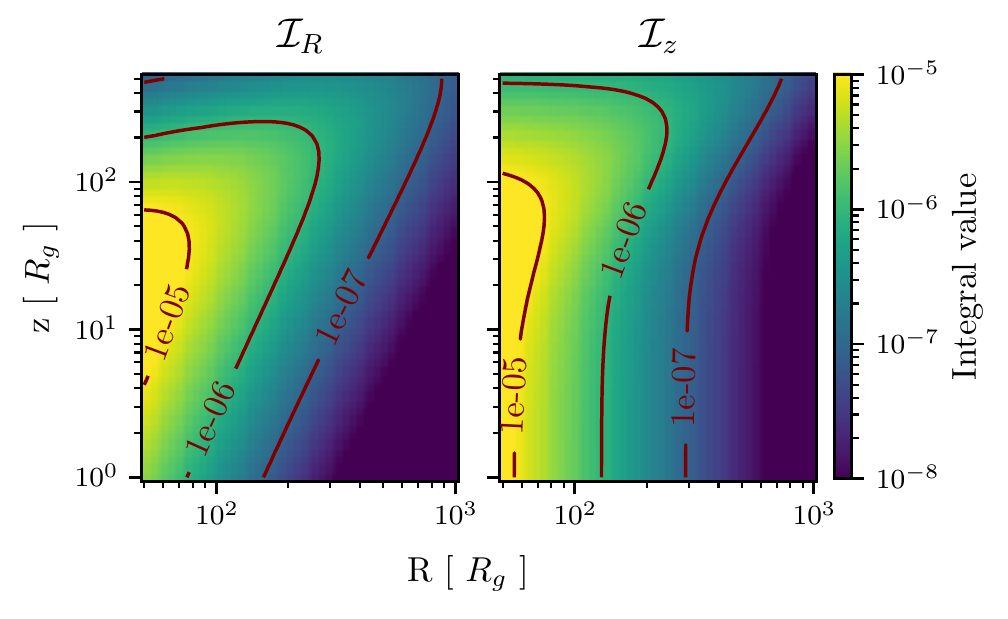}
    \includegraphics[width=\columnwidth]{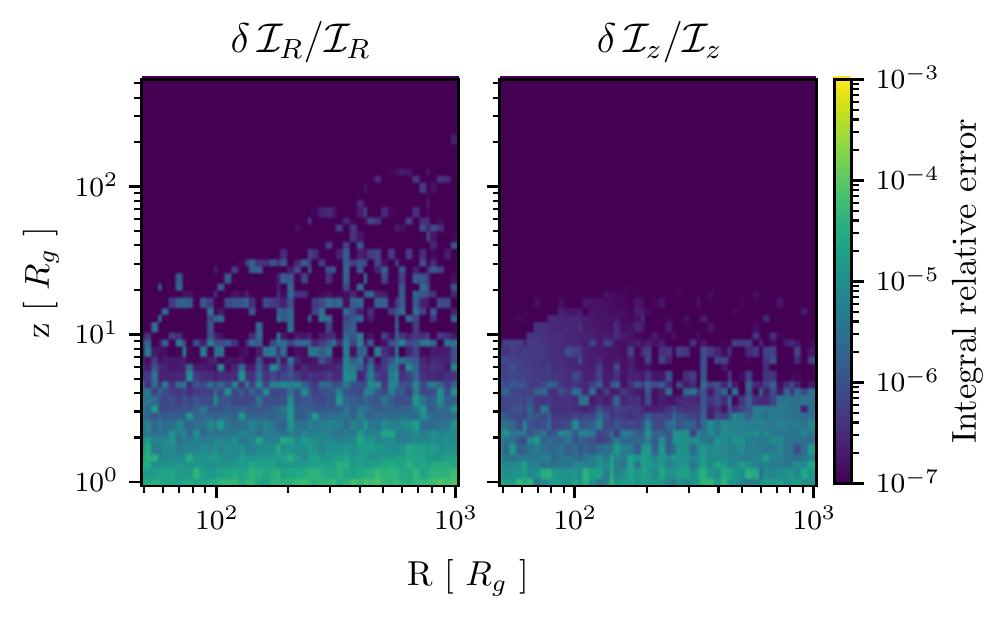}
    \caption{Top panels: Values for the radial and height integrals across the $R-z$ grid. Bottom panels: Relative error of the integrals. Note that the relative error stays well below $10^{-3}$ for the whole variable range, the low height points being the most difficult to compute.}
    \label{fig:integral_convergence}
\end{figure}

Numerically solving the integrals (\ref{eq:Ir}) and (\ref{eq:Iz}) can be tricky because the points $\Delta = 0$ are singular. We use the \textsc{Quad} integration method implemented in the \textsc{Scipy} \citep{virtanen_scipy_2019} Python package to compute them. We fix the absolute tolerance to 0, and the relative tolerance to $10^{-4}$, which means the integral computation stops once it has reached a relative error of $10^{-4}$. We have checked that the integrals converge correctly by evaluating the integration error over the whole grid, as can be seen in Figure \ref{fig:integral_convergence}. The relative errors stays below $10^{-3}$, which is 10 times more the requested tolerance but still a good enough relative error. We thus set a tolerance of $10^{-4}$ as the code's default.

\subsection{Solver convergence}

\begin{figure}
    \centering
    \includegraphics[width=\columnwidth]{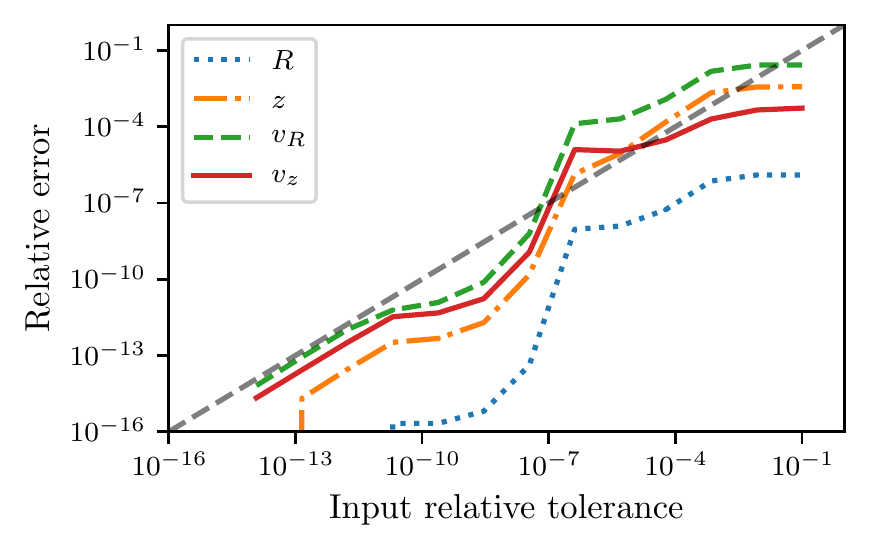}
    \caption{Measured relative error as a function of the input relative tolerance for the \textsc{Assimulo} IDA solver. The black dashed line represents equality. }
    \label{fig:solver_error}
\end{figure}

To assess the convergence of the \textsc{IDA} solver, we calculate the same gas trajectory multiple times changing the input relative tolerance of the solver, from $10^{-15}$ to $10^{-1}$. We take the result with the lowest tolerance as the true value, and compute the errors of the computed quantities, $R, z, \varv_R, \varv_Z$ relative to our defined true values. As we can see in Fig. \ref{fig:solver_error}, the relative error is well behaved and generally accomplishes the desired tolerance. After this assessment we fix the relative tolerance to $10^{-4}$ as the code's default.

{}

\bsp	
\label{lastpage}
\end{document}